\documentclass[useAMS,usenatbib,usegraphicx]{mn2e}
\usepackage{amssymb}

\title[Dark matter dominated dSph stellar density profiles]{The shapes of dark matter
dominated dwarf spheroidal stellar density profiles
}
\date{Accepted 2002 May 27. Received 2002 March 25; in original form 2001 July 13}

\pagerange{\pageref{firstpage}--\pageref{lastpage}} \pubyear{2002}

\author[K. Sigurdson and N. Turok]{Kris Sigurdson\thanks{Email:
sigurdson@canada.com, ksigurds@caltech.edu (KS)} and Neil Turok\thanks{Email:
N.G.Turok@damtp.cam.ac.uk (NT)}\\
DAMTP, Centre for Mathematical Sciences, Wilberforce Road, Cambridge,
CB3 0WA, UK
}

\begin{document}

\label{firstpage}

\maketitle

\begin{abstract}

A Bayesian likelihood analysis of the stellar density profile shapes of
several dark matter dominated Galactic dwarf spheroidal galaxies (dSphs) has been completed.  This
analysis indicates that the inner power law, $\rho \propto
r^{-\gamma}$, of each of these profiles satisfies
$\gamma < 1.21$ at greater than 99\% confidence with a slight preference
for
{$\gamma \sim 0.7$.}  Speculating that this bound on the stellar
profiles may be related to the inner
power law of the underlying dark matter distribution reveals a statistical
discrepancy with the values favoured by recent N-body
simulations of cold dark matter (CDM), $\gamma_{\scriptscriptstyle CDM} \sim 1.4-1.5$.

\end{abstract}


\section{Introduction}

The nature of the dark matter remains one of the prime areas of
astrophysical and cosmological research.  There has
been some controversy recently concerning cold collisionless dark
matter (CDM), arguably the simplest solution to dark matter problem.
Assuming that satellite dark haloes can be associated with dwarf
spheroidal galaxies (dSphs), CDM appears to overpredict by two orders of
magnitude the
number of these satellite haloes surrounding the Galactic halo
\citep{moore1999a}.  Recent N-body simulations of CDM also indicate that
dark matter haloes
have cuspy cores on subgalactic scales,
with an inner exponent $\gamma_{\scriptscriptstyle CDM} \sim 1.4 - 1.5$
\citep{fukus1997,moore1999b,jing2000,subra2000,bullo2001}, up from the
value
found in earlier simulations of $\gamma_{\scriptscriptstyle CDM} \sim
1.0$ \citep[][hereafter NFW]{nfw1996}.

Recent detailed, high resolution observations of the rotational motion
of LSB galaxies indicate their core density profiles are substantially
flatter than CDM predictions (\citealt{blok2001}; see also
\citealt{borri2001}).  (For earlier
discussion and debate, see for example \citet{bosch2000} and
\citet{moore1999b}).

Since the discrepancy between CDM theory and observation is greatest on
small scales, it is natural to seek to sharpen it further by examining
the smallest, nearest galaxies, namely dwarf spheroidal galaxies.
Galactic dSphs have long been recognised as excellent probes of the
nature of the dark matter.  Besides being the smallest known galaxies
they also possess the highest known ${M}/{L}$ ratios.  Observationally,
their proximity means that individual stars can be resolved, so that
beam smearing effects of the type discussed by \citet{bosch2000} for
LSBs are not an issue.  Unfortunately, whilst velocity measurements
have established the presence of large amounts of dark matter, they
have hitherto not been sufficiently accurate to determine the radial
profile of the dark matter density \citep{arman1997}.

However, the stellar density profile is rather well established
\citep{irwin1995} and enough data exists for a detailed comparison
with the theoretically predicted dark matter profiles. 
This comparison is only meaningful if the stars trace the mass, a very strong
assumption, which is hard to justify.  Nevertheless, 
it seems a reasonable first step to fit dSph stellar
profiles to the parametric form used in theoretical studies of dark matter 
profiles, whilst
acknowledging that any conclusion on the dSph dark matter profile must
await more accurate velocity measurements.  As both theory and
observations develop, we anticipate that a more complete likelihood
analysis, of the type performed here but including both density and
velocity data, should become possible. 

One should also point out that at least for some star formation histories,
comparing the stellar and dark matter profiles is not without meaning. 
If the bulk of the stars formed before the dSph 
dark matter haloes,
they would thereafter 
cluster under gravity just like dark matter particles. 
If, on the other hand, stars formed after the dark matter haloes, from
baryons which were heated when the Universe was reionised at $z
\ga 5$, or alternatively if the stars in dSphs were accreted from
other parts of the galaxy, there would not necessarily be any simple
relation between the stellar, and dark matter density profiles.

CDM theory over-predicts both the number density, and the velocity
dispersion of haloes with the masses of dwarf spheroidals
\citep{bode2000}.  If CDM is to be viable, there must
exist some mechanism (such as stellar feedback)
to suppress star formation
in most, but not all, low mass haloes. The surviving low mass galaxies
would then be those in which stars formed earliest, and those which
possessed the
highest central density. These selection effects would tend to
favor haloes with steeper-than-typical inner profiles  and
higher central densities. 
Comparing a typical CDM halo profile
against observed dark matter profiles would 
actually be conservative 
in this respect. 

In a similar way we may also speculate on how compatible other
theories of dark matter are with observations of dark matter dominated
dwarf spheroidals.  For example, it has been suggested that a warm dark matter (WDM) scenario may better match the observed
number of satellite galaxies predicted for the local group
\citep{colin2001} and so a mechanism to suppress star formation may be
less necessary.  Additionally, stars may be a more viable tracer of the
dark matter in WDM theories as the recollapse of small WDM haloes is
expected to occur considerably later than in the CDM case.  However,
despite these advantages, theories of WDM may predict haloes cuspy
enough to be cause for concern as in the CDM case \citep{eke2001}.  
 
\section{Methodology}

We have studied the radial distribution of stellar matter in the dark
matter dominated Galactic dSphs using the framework of Bayesian
likelihood analysis \citep{grego1992}.

The stellar density profile was assumed to take the smoothly
interpolating double power law form of equation (1)
\citep[e.g.,][]{kravt1998}.

\begin{equation}
\rho(r)=\frac{\rho_{0}}{
({r}/{r_0})^{\gamma}[0.5+0.5({r}/{r_0})^{\alpha}]^{({\beta-\gamma})/{\alpha}}}
\end{equation}

Using this form yields $\rho \propto r^{-\gamma}$ for radial distances
smaller than $r_0$, and $\rho \propto r^{-\beta}$ for radial distances
larger than $r_0$.
The parameter $\alpha$ controls the scale over which the power law
shifts from $\gamma$ to $\beta$.  Roughly speaking, this change occurs
between radii $e^{- 1/\alpha}r_0$ and $e^{1/\alpha}r_0$.
This density profile
accommodates a wide range of useful models, including a modified
King-like profile
\citep{king,rood1972} for $(\alpha,\beta,\gamma)=(2,3,0)$, the NFW profile
for
$(\alpha,\beta,\gamma)=(1,3,1)$, and the profile suggested by recent
N-body CDM simulations $(\alpha,\beta,\gamma)=(1,3,1.5)$.

Projecting onto the celestial sphere gives the effective two
dimensional stellar surface density.

\begin{equation}
\Sigma(r_{\bot})=2\int_{0}^{\infty}\rho\Big(\sqrt{r_{\bot}^2+z^2}\Big) \
dz
\end{equation}

Currently, the best observational determination of stellar density as a
function of
radius in the dSphs of interest is the work of \citet{irwin1995}, and  can
be summarised
in this analysis by a vector ${{\widetilde{\Sigma}_{obs}}}$ which contains
the
mean density at a discrete set of radii and an associated covariance
matrix ${{\widetilde{C}_{obs}}}$.

For each model $(\alpha,\beta,\gamma,r_0,\rho_0)$ we have calculated the
conditional relative likelihood
$\mathcal{L}(\alpha,\beta,\gamma,r_0,\rho_0
| {{\widetilde{\Sigma}_{obs}}})$ --  a $\chi^2$ statistic in practice --
to
quantify the compatibility of a particular
model with observations.

The chief advantage of the Bayesian formalism is that it allows one to
assess the relative likelihood of a particular subclass of the models
by integrating $\mathcal{L}$ over the relevant equiclassification
surfaces
in the model space.  To assess the relative
likelihood of different values of $\gamma$ we integrate $\mathcal{L}$
over all models with the same $\gamma$ but different
$(\alpha,\beta,r_0,\rho_0)$ and obtain the marginalized distribution
$\mathcal{L}_{\gamma}(\gamma | {{\widetilde{\Sigma}_{obs}}})$.  This
process
allows for a fair decoupling of our state of knowledge of $\gamma$
from $(\alpha,\beta,r_0,\rho_0)$ without, for example, ignoring the data
in the
outer regions of the dSph in an ad hoc manner.

Care must be taken when choosing the ranges of parameter values to
ensure that a fair sampling of the parameter space is made into
all regions that can contribute significantly to the marginalized
distribution.  The reason for this is the edges of the parameter
distribution impose the trivial prior that the parameters of candidate
models must lie
within the range of parameters chosen.  For the purposes of this study
we have found that a largely unbiased range of parameters is:
{$\alpha \in [0.5,12]$; \hspace{0.3em} $\beta
\in [2.5,17]$;
\hspace{0.3em} $\gamma \in [-13,3]$;   \hspace{0.3em} $r_0 \in
\mathbb{R}$;  and  \hspace{0.3em}$\rho_0 \in \mathbb{R} $.
}

  In practice $\mathcal{L}$ is an extremely rapidly decreasing function of
$(r_0,\rho_0)$ about its maximum, and so for computational purposes the
marginalization
over these variables has been approximated by a
maximization.   We find that maximization over $(r_0,\rho_0)$ yields
indistinguishable results to marginalization provided that the
characteristic turnover radius, $r_0$, is not permitted to be
unreasonably large (ie. larger than the edge of the galaxy).
Marginalization over $(\alpha,\beta)$ proceeds
straightforwardly to yield $\mathcal{L}_{\gamma}$.

The three dSphs we have focused our analysis on, Carina, Draco and
Ursa Minor have mean ellipticities,~$\epsilon = 1- {b}/{a}$, of
$0.33$, $0.29$, and $0.56$ respectively and the available surface
density data is averaged over elliptical bins.  To make a conservative
analysis in spherical theory we made a direct comparison of
circularly rebinned surface density data with the theoretical
profiles, using the properly reweighted error distribution.
In general, we find that even the rough approximation of comparing
the elliptically binned density profile with spherical theory gives
similar limits on $\gamma$ as the gross dependence of $\mathcal{L}$
on the shape parameters is insensitive to the exact data binning scheme.
This is expected as spherically smoothing an
ellipsoidal distribution has the effect of smoothing the radial
distribution
over scales of $\sim \epsilon a$ without significantly altering the gross
radial dependence -- especially in the nearly flat inner regions
important for this analysis.

\section{Discussion}

The marginalized likelihood distribution $\mathcal{L}_{\gamma}$ for
the three dSphs in question is shown in Figure 1.  Although the peak of
the distribution fluctuates slightly from galaxy to galaxy the upper
bound on $\gamma$ is consistent between all three.

\begin{figure}
\includegraphics*[width=3.33in]{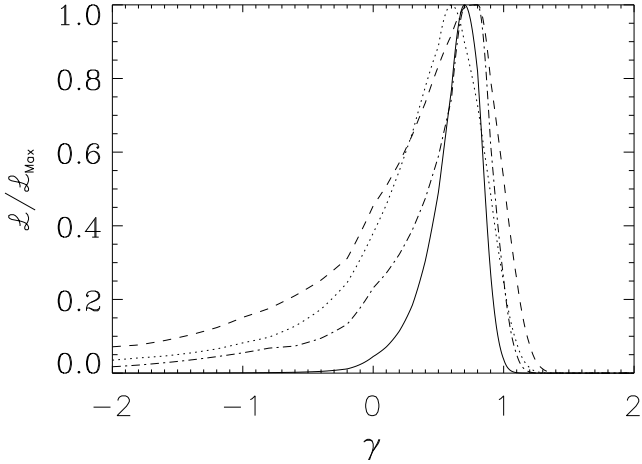}
\caption{The likelihood distribution of $\gamma$, the exponent of the
inner
power law, for Carina (dot-dashed), Draco (dotted), Ursa Minor (dashed),
and for the three galaxies taken together as an ensemble (solid).}
\end{figure}

While the distributions in $\gamma$ have tails stretching well
into the negative region of parameter space this is not
unexpected.
Negative values of $\gamma$ are simply models that have a peak in
the density at a scale of $\sim r_0$, a plausible result given the
adopted uncertainties.  Projection effects coupled with the freedom
to renormalize and rescale the density profile have the result of
extending the tail relatively far into the
negative $\gamma$ region.  Despite the ambiguity between models in the
negative $\gamma$ regime, it is clear from the Figure 1. that models with
$\gamma > 1.21$ are strongly excluded by this analysis.

The best-fitting density profiles are shown in Figure 2.  Note that while
the profiles are nearly always good fits for the inner and central
regions there sometimes appears to be an excess of density for the
outer few data points -- related to the scheme adopted for background
subtraction of stars \citep{irwin1995}.  The results with respect to the
parameter
$\gamma$ do not change if these points are included or excluded from
the analysis.  We have adopted the strategy of including all available
data.

\begin{figure}
\includegraphics*[width=3.33in]{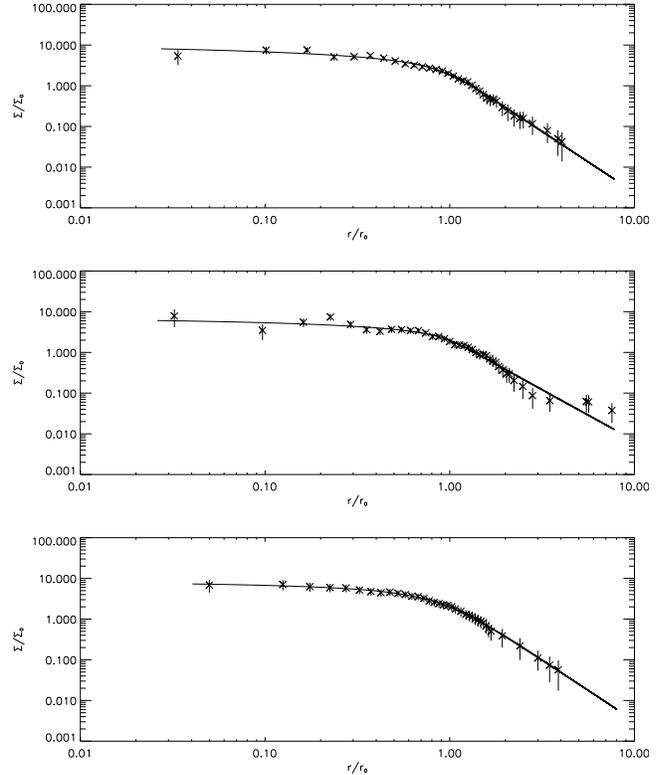}
\caption{Shown are the maximum likelihood models
$(\alpha,\beta,\gamma)=(5.5,4,0.7)$ for Carina (Top),
$(\alpha,\beta,\gamma)=(9.5,3.5
,0.6)$ for Draco (Middle), and
$(\alpha,\beta,\gamma)=(3.5,4.0,0.5)$ for Ursa Minor (Bottom).
Models with less sharp turnovers of $\alpha \approx 1 - 2$ can also be
found that yield reasonable fits.}
\end{figure}

Table 1 summarises the limits this analysis places on $\gamma$ and the
other shape parameters for each galaxy on its own, and the galaxies
considered together as an ensemble.  It also presents the limits with a
tophat prior of $1 \le \alpha \le 2$ imposed on the likelihoods --
the range of parameter space that the NFW and recent N-body CDM
profiles are contained in. The effect of this prior is to shift the
peak of the distribution and the upper
bound on $\gamma$ to even \emph{lower} values than when the full range
of $\alpha$ is used.

\begin{table}
\caption{Summary of shape parameter limits. \label{tbl-1}}
\label{symbols}

\begin{tabular}{lcccc}

\hline
\multicolumn{5}{c}{Uniform Priors} \\
\hline

Galaxy     & $\gamma (99\%)^{*}$& $\gamma_{peak}$ & $\beta$ $(2
\sigma)$ & $\alpha$ $(2 \sigma)$ \\

\hline

Carina     & $<$$1.09$ & $0.80$ & $4.0^{+2.3}_{-0.6}$
& $<$$5.2$ \\
Draco      & $<$$1.12$ & $0.60$ & $3.5^{+1.9}_{-0.3}$
& $<$$7.3$ \\
Ursa Minor & $<$$1.21$ & $0.80$ & $4.0^{+4.5}_{-0.8}$
& $<$$2.7$ \\

Ensemble   & $<$$0.98$ & $0.70$ & $4.0^{+0.8}_{-0.5}$
& - \\

\hline
\multicolumn{5}{c}{$1 \le \alpha \le 2$ Tophat Prior} \\
\hline

Carina     & $<$$0.76$ & $0.00$ & $5.0^{+4.9}_{-1.1}$
& - \\
Draco      & $<$$0.76$ & $0.00$ & $5.0^{+5.0}_{-1.4}$
& - \\
Ursa Minor & $<$$0.95$ & $0.00$ & $5.0^{+6.8}_{-1.5}$
& - \\

Ensemble   & $<$$0.52$ & $0.00$ & $5.0^{+2.8}_{-0.8}$
& - \\

\hline

\end{tabular}

\medskip

{*}{These values are based on the one sided Gaussian
positive deviation from the peak.  Calculating with the full two
sided distribution yields an even stronger upper confidence bound on
$\gamma$.}

\end{table}

As measurements of the radial profile of the velocity dispersion in
these galaxies increase in sensitivity a joint analysis fitting both
the stellar dispersion and stellar density profiles predicted by a
given DM profile should provide a much better test of the CDM and
similar paradigms.

\section{Conclusion}

We have completed a Bayesian likelihood analysis of the stellar
density profile shapes of the Galactic dwarf spheroidal galaxies Draco, Ursa Minor, and Carina.
With uniform prior assumptions for the parameters
$(\alpha,\beta,\gamma,r_0,\rho_0)$ we conclude that, for each galaxy,
$\gamma < 1.21$ at greater than 99\% confidence.

As discussed in the introduction, except in the special situation
where the bulk of the stars formed before the dark matter halo,
$\gamma_{\scriptscriptstyle DM}$ would not be expected to be the same
as
$\gamma{\scriptscriptstyle stars}$.  Nevertheless the comparison we
made quantifies the the discrepancy between the observed inner stellar
profile of dSphs and simulated dark matter haloes, a discrepancy which
CDM theory or one of its cousins must ultimately explain.

\section*{Acknowledgements}

We would like to thank M. Irwin and D. Hatzidimitriou for kindly
providing their data for use in this analysis.  KS was supported in
part by funds from the Government of British Columbia (Canada) through
the 2000 Queen Elizabeth II British Columbia Centennial scholarship, and
the generosity of the Cambridge Commonwealth Trust.  NT is supported
by PPARC.

\label{lastpage}


\begin{thebibliography}{}

\bibitem[Armandroff, Pryor \& Olszewski (1997)]{arman1997} Armandroff T.E., Pryor C., \&
Olszewski E., 1997, IAUJD, 2E, 35A

\bibitem[Avila-Reese et al.(2001)]{colin2001} Avila-Reese V., Colín P.,
Valenzuela O., D'Onghia E., Firmani, C., 2001, ApJ, 559, 516

\bibitem[Borriello \& Salucci(2001)]{borri2001} Borriello A., \& Salucci
P., 2001, MNRAS, 323, 285

\bibitem[Bode, Ostriker, \& Turok(2000)]{bode2000} Bode P., Ostriker J.P.,
 \& Turok N., 2000, AAS, 197, 72.04

\bibitem[de Blok et al.(2001)]{blok2001} de Blok W.J.G., McGaugh S.S.,
Bosma A., \& Rubin V.C., 2001, ApJ, 552, L23

\bibitem[Bullock et al.(2001)]{bullo2001} Bullock J.S., Kolatt T.S.,
Sigad Y., Somerville R.S., Kravtsov A.V., Klypin A.A., Primack J.R., \&
Dekel A., 2001, MNRAS, 321, 559

\bibitem[Eke et al.(2001)]{eke2001} Eke V.R., Navarro J.F., Steinmetz
M., 2001, ApJ, 554, 114

\bibitem[Fukushige \& Makino(1997)]{fukus1997} Fukushige T., \& Makino J.,
1997, ApJ, 477, L9

\bibitem[Gregory \& Loredo(1992)]{grego1992} Gregory P.C., \& Loredo T.J.,
1992, ApJ, 536, L63

\bibitem[Irwin \& Hatzidimitriou(1995)]{irwin1995} Irwin M., \&
Hatzidimitriou D., 1995, MNRAS, 277, 1354

\bibitem[Jing \& Suto(2000)]{jing2000} Jing Y.P., \& Suto Y., 2000, ApJ,
529, L69

\bibitem[King(1962)]{king} King I.R., 1962, ApJ, 67, 471

\bibitem[Kravtsov et al.(1998)]{kravt1998} Kravtsov A.V., Klypin
A.A., Bullock J.S., \& Primack J.R., 1998, ApJ, 502, 48

\bibitem[Mateo(1997)]{mateo1997} Mateo M., 1997, in {\it The Nature of
Elliptical Galaxies} Vol. 116, ed. M. Arnaboldi, G.S. Da Costa,
P. Saha, ASP, San Francisco, p. 259-269

\bibitem[Mateo(1998)]{mateo1998} Mateo M., 1998, ARAA, 36, 435

\bibitem[Moore et al.(1999a)]{moore1999a} Moore B., Ghigna S.,
Governato F., Lake G., Quinn T., Stadel J., \& Tozzi P., 1999, ApJ, 524,
L19

\bibitem[Moore et al.(1999b)]{moore1999b} Moore B., Quinn T., Governato
F., Stadel J., \& Lake G., 1999, MNRAS, 310, 1147

\bibitem[Navarro, Frenk, \& White(1996)]{nfw1996} Navarro J.F., Frenk
C.S., White S.D.M., 1996, ApJ, 462, 563

\bibitem[Rood et al.(1972)]{rood1972} Rood H.J., Page T.L., Kintner E.C.,
\& King I.R., 1972, ApJ, 175, 627

\bibitem[Subramanian et al.(2000)]{subra2000} Subramanian K., Cen R., \&
Ostriker J.P., 2000, ApJ, 538, 528

\bibitem[van den Bosch \& Swaters(2000)]{bosch2000} van den Bosch F.C., \&
Swaters R.A., 2000, MNRAS, 325, 1017

\end{thebibliography}
\end{document}